\documentclass[preprint,12pt]{elsarticle}

\usepackage[retain-explicit-plus]{siunitx}
\usepackage{makecell}
\usepackage{lipsum}
\usepackage{rotating}
\usepackage{lineno}
\usepackage{hyperref}
\usepackage{mathtools}
\usepackage{multirow}
\usepackage{multicol}
\usepackage{booktabs}
\usepackage{xcolor}
\usepackage{gensymb}
\usepackage{subcaption}

\usepackage{listings}
\usepackage{amssymb}
\usepackage{pifont}

\newcommand{\xmark}{\ding{55}}

\usepackage{cleveref}
\journal{Expert Systems with Applications}
\biboptions{authoryear}

\usepackage{blindtext}
\usepackage{hyperref}
\usepackage{nameref}

\newcounter{mylabelcounter}

\makeatletter
\newcommand{\labelText}[2]{%
\refstepcounter{mylabelcounter}%
\immediate\write\@auxout{%
 \string\newlabel{#2}{{\unexpanded{#1}}{\thepage}{{\unexpanded{#1}}}{mylabelcounter.\number\value{mylabelcounter}}{}}%
}%
}
\makeatother

\begin{document}

\begin{frontmatter}
\lstset{basicstyle=\normalsize\ttfamily,breaklines=true}

\title{Explainable assessment of financial experts' credibility by classifying social media forecasts and checking the predictions with actual market data}

\author[mymainaddress]{Silvia García-Méndez\corref{mycorrespondingauthor}}
\ead{sgarcia@git.uvigo.es}
\author[mymainaddress]{Francisco de Arriba-Pérez}
\ead{farriba@gti.uvigo.es}
\author[mymainaddress]{Jaime González-González}
\ead{jaimegonzalez@gti.uvigo.es}
\author[mymainaddress]{Francisco J. González-Castaño}
\ead{javier@det.uvigo.es}
\address[mymainaddress]{Information Technologies Group, atlanTTic, University of Vigo, Vigo, Spain}

\cortext[mycorrespondingauthor]{Corresponding author: sgarcia@gti.uvigo.es}

\begin{abstract}
Social media include diverse interaction metrics related to user popularity, the most evident example being the number of user followers. The latter has raised concerns about the credibility of the posts by the most popular creators. However, most existing approaches to assess credibility in social media strictly consider this problem a binary classification, often based on a priori information, without checking if actual real-world facts back the users' comments. In addition, they do not provide automatic explanations of their predictions to foster their trustworthiness. In this work, we propose a credibility assessment solution for financial creators in social media that combines Natural Language Processing and Machine Learning. The reputation of the contributors is assessed by automatically classifying their forecasts on asset values by type and verifying these predictions with actual market data to approximate their probability of success. The outcome of this verification is a continuous credibility score instead of a binary result, an entirely novel contribution by this work. Moreover, social media metrics (\textit{i.e.}, user context) are exploited by calculating their correlation with the credibility rankings, providing insights on the interest of the end-users in financial posts and their forecasts (\textit{i.e.}, drop or rise). Finally, the system provides natural language explanations of its decisions based on a model-agnostic analysis of relevant features.
\end{abstract}


\end{frontmatter}

\section{Introduction}
\label{sec:introduction}

Data credibility depends on the information source (\textit{e.g.}, humans or bots), the content, and its context (\textit{e.g.}, user relations, content domain, etc.). This problem has gained renewed attention owing to the popularity of social media \citep{Kaliyar2021,Nian2021} and their anonymity and effortless fast information-sharing features \citep{Huang2021}. Social media posts and user profiles are valuable input data to assess creators' credibility for decision-making \citep{Kurniati2017}. For example, identifying reliable and trustworthy creators is crucial to fight disinformation from fake news (\textit{i.e.}, deliberate disinformation) \citep{kozik2022technical,Verma2022}. Nevertheless, social media platforms lack, in general, appropriate mechanisms to distinguish valuable from useless data \citep{Kamkarhaghighi2016,Hanaforoosh2021}, hence the interest in these mechanisms \citep{Huang2021,Jing2021}.

Most existing research on user credibility has focused on analyzing shared content, rather than considering the context of the creators and their profiles \citep{Asfand2023,li2023online}. We can identify different approaches: (\textit{i}) graph-based (\textit{i.e.}, measuring the users' credibility from other directly or indirectly connected users) \citep{Qureshi2021}, (\textit{ii}) those performing Sentiment Analysis (\textsc{sa}) of the content \citep{Kurniati2017}, and (\textit{iii}) solutions with domain-dependent features \citep{Huang2019}. Additional features related to the users' profiles (\textit{i.e.}, the number of followers) and their demographic data (\textit{e.g.}, age) have been identified as helpful \citep{Verma2022}. As an example of a complete approach combining diverse strategies, the recent work by \citet{li2023online} describes a Machine Learning (\textsc{ml}) based assessment tool of Twitter profiles applying \textsc{sa}, content-based features and user-context features. It is also interesting to observe that most current solutions for credibility analysis tend to approach the problem as a strictly binary classification task, instead of producing continuous rankings \citep{Zhou2017}.

Moreover, eXplainable Artificial Intelligence (\textsc{xai}) techniques are scarce in the literature on the target problem. These techniques take advantage of the intrinsically interpretable nature of some \textsc{ml} models, or are \textit{ad hoc} solutions to circumvent the opacity of black-box algorithms \citep{Jung2022,akbar2023trustworthy}. They help professionals and end-users better understand the decisions of the \textsc{ml} models to foster their trustworthiness. The most popular \textsc{xai} approaches are (\textit{i}) counterfactual descriptions \citep{Chen2023}, (\textit{ii}) feature importance \citep{Huang2021}, (\textit{iii}) natural language explanations \citep{Cambria2023}, and (\textit{iv}) visual dashboards \citep{Batic2024}.

Summing up, our work focuses on the financial domain, owing to the many credibility concerns about stock market forecasts \citep{Huang2019}. It contributes to the investors' well-informed decision-making through reliable and updated information from social media \citep{Evans2021}. Specifically, we classify financial posts in forecast categories and automatically verify if actual market events back these predictions to measure the level of success of the users. This success percentage is presented as a continuous credibility score instead of a binary result. Additionally, user context is exploited by comparing credibility scores with social media user metrics, providing insights related to the user interests about the forecasts in financial posts. Finally, the model predictions are automatically explained in natural language.

The rest of this paper is organized as follows. Section \ref{sec:related_work} reviews the relevant works on user credibility assessment. Section \ref{sec:methodology} introduces the proposed solution. Section \ref{sec:results} describes the experimental data set, the implementations, and the results obtained. Finally, Section \ref{sec:conclusions} highlights the contributions and proposes future research. 

\section{Related work}
\label{sec:related_work}

\subsection{Social media and \textsc{ml} for credibility assessment}

Since the information in social media may come from untrustworthy users, filtering advertisements, scams, and spam from credible data is crucial in professional domains \citep{Huang2019}. 

There exists plenty of prior work on social media credibility assessment \citep{Huang2019}, covering domains such as business \citep{Abdullah2021,Al-Yazidi2022}, disaster scenarios \citep{Assery2022}, news \citep{Pelau2023,Shrestha2024}, politics \citep{Page2018} and finance \citep{Evans2021}.

Two typical limitations of current works on credibility assessment are that (\textit{i}) they often approach the problem as a binary classification in which user credibility is not ranked as a continuous variable; and (\textit{ii}) they do not provide explanations on their predictions. Also, many works ignore the user's context and focus on the content. The latter is not the case, for example, in the work by \citet{Prada2020}, who presented a user reputation prediction solution for Wallapop based on generalized linear models (\textsc{glm}s) (parametric models that are more straightforward than non-parametric traditional \textsc{ml} algorithms). The features included elementary linguistic ones, the bad word ratio, and the user's Twitter profile data.

Earlier works by \citet{Zhou2017,Abu-Salih2019} assessed the credibility of diverse sources based on rules. The analysis by \citet{Alrubaian2017} relied on the relations in a network of reputable individuals to detect and prevent malicious events in social networks. \citet{Bukhari2017} classified tweets into high- and low-impact ones by analyzing their content with \textsc{sa}. They also predicted the users' areas of expertise by considering some user-context features.

\subsection{Social media and \textsc{ml} for financial credibility assessment}

In the financial domain, we should mention the work by \citet{Kamkarhaghighi2016}, which measured credibility from the number of seed tweets of a user, the number of stock market-oriented followers identified with \$cashtags, and the ratio between these and other followers. Conversely, \citet{Huang2019} calculated trust scores based on PageRank (\textsc{pr}) and Hyperlink-Induced Topic Search (\textsc{hits}) models to identify credible data in financial tweets. A list of credible users was extracted from prestigious independent financial authorities. Four trust filters were proposed: experience, expertise, authority, and reputation. The first two considered the user posting frequency, while the others were based on the user's social network structure, favoring well-connected users. Then, the authors applied \textsc{sa} to the tweet content. \citet{Evans2021} followed a supervised classification approach trained with metadata from the content and the user context, obtaining the best results with tree-based models. They employed counters of Bag of Words (\textsc{bow}), Part of Speech (\textsc{pos}) tags, and \textsc{url}s. Moreover, \citet{Huang2021} proposed a solution to identify credible Twitter users based on trustworthy experts. As \citet{Kamkarhaghighi2016}, they filtered finance-oriented users with \$cashtags and took financial experts from well-known sources. Then, they trained an \textsc{ml} model with content-oriented and user context features. Interpretability was provided by checking the importance of the features of a Random Forest (\textsc{rf}) classifier.

\subsection{Contribution}

In light of the comparison with competing solutions in Table \ref{tab:comparison}, our work is the first that (\textit{i}) assesses user credibility from social media financial posts that are automatically classified into types of forecasts and then validated with actual market data, (\textit{ii}) provides a credibility ranking as a continuous scale, (\textit{iii}) analyzes the correlations between this credibility ranking and diverse user context metrics to gain insights on the interest of the audience in financial posts, and (\textit{iv}) describes the model predictions with natural language explainability techniques. Our solution has been validated using financial posts, and unlike other \textsc{xai}-based systems in the literature, it exploits the relevance of the features to create natural language descriptions. Even though the training process is time-consuming compared to rule-based approaches presented by \citet{Kamkarhaghighi2016} and \citet{Huang2021}, and the approach based on \textsc{pr} and \textsc{hits} by \citet{Huang2019}, the prediction is performed in real-time. In contrast to \citet{Evans2021} and \citet{Huang2021}, it offers a credibility ranking on a continuous scale.

\begin{table}[!htbp]
\centering
\footnotesize
\caption{Comparison of financial user credibility solutions considering the approach, type of features used (content and user context), and availability of continuous-scale ranking (Rank), real-world validation (Val.), and explainability (Ex.)—including feature relevance (\textsc{fr}) and natural language (\textsc{nl}).}
\label{tab:comparison}
\begin{tabular}{lcccccc}
\toprule
\textbf{Authorship} & \textbf{Social} & \textbf{Approach} & 
\textbf{Features} &
\textbf{Rank} & 
\textbf{Val.} &
\textbf{Ex.}\\
& \textbf{media}\\
\midrule

\multirow{2}{*}{\citet{Kamkarhaghighi2016}} & \multirow{2}{*}{Twitter} & Rules & \multirow{2}{*}{User} &\multirow{2}{*}{\checkmark} & \multirow{2}{*}{\xmark} & \multirow{2}{*}{\xmark}\\
& & \textsc{ml}\\\midrule

\multirow{2}{*}{\citet{Huang2019}} & \multirow{2}{*}{Twitter} & \multirow{2}{*}{\textsc{pr}, \textsc{hits}} & Content & \multirow{2}{*}{\checkmark} & \multirow{2}{*}{\xmark} & \multirow{2}{*}{\xmark}\\
& & & User\\\midrule

\multirow{2}{*}{\citet{Evans2021}} & \multirow{2}{*}{Twitter} & \multirow{2}{*}{\textsc{ml}} & Content & \multirow{2}{*}{\xmark} & \multirow{2}{*}{\xmark} & \multirow{2}{*}{\xmark}\\
& & & User\\\midrule

\multirow{2}{*}{\citet{Huang2021}} & \multirow{2}{*}{Twitter} & Rules & Content & \multirow{2}{*}{\xmark} & \multirow{2}{*}{\xmark} & \multirow{2}{*}{\textsc{fr}}\\
& & \textsc{ml} & User\\\midrule\midrule

\multirow{2}{*}{\textbf{Proposed solution}} & Twitter & \multirow{2}{*}{\textsc{ml}} & \multirow{2}{*}{\makecell{Content \\ User}} & \multirow{2}{*}{\checkmark} & \multirow{2}{*}{\checkmark} &\textsc{fr} \\
& Yahoo & & & & &\textsc{nl}\\

\bottomrule
\end{tabular}
\end{table}

\section{Methodology}
\label{sec:methodology}

\begin{figure}[!htpb]
 \centering
 \includegraphics[scale=0.18]{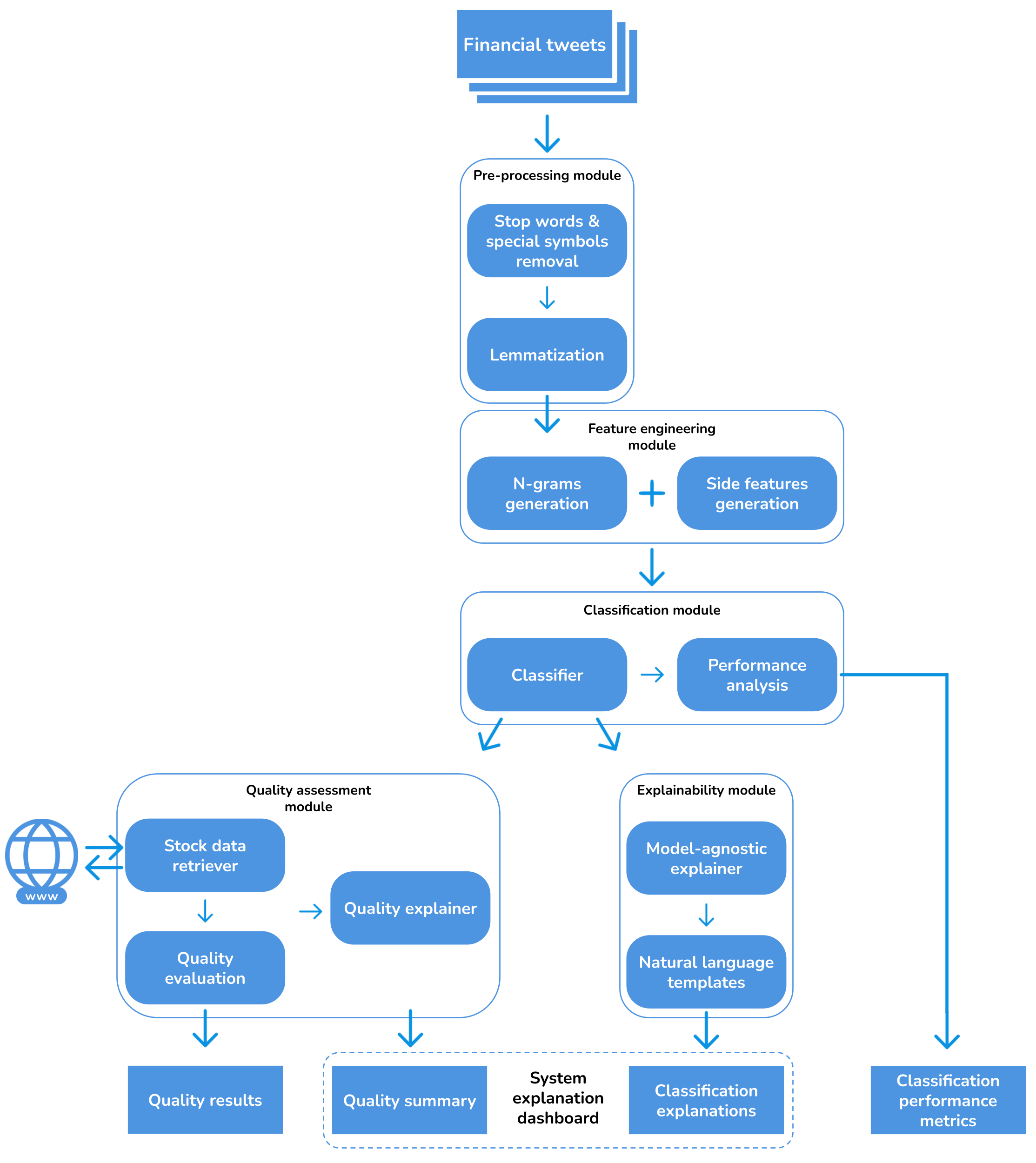}
 \caption{System scheme.}
 \label{fig:scheme}
\end{figure}

Figure \ref{fig:scheme} shows the scheme of the proposed solution for the credibility assessment of finance-oriented users. Firstly, the pre-processing module prepares the textual content. Then, the feature engineering module generates $n$-grams and side features. The hybrid classification module automatically classifies posts in types of forecasts with term detectors and \textsc{ml} algorithms. Prediction quality is ranked by the quality assessment module and described by the explainability module.

\subsection{Pre-processing module}
\label{sec:preprocessing}

Prior to feature extraction is essential to remove meaningless data to ensure optimal knowledge extraction from the textual content. Regular expressions are applied to remove links to external sources (\textit{e.g.}, journals, blogs, etc.) and images. Stop words and special characters (\textit{e.g.}, currency symbols, percentage symbols, etc.) and emoticons are discarded. Finally, the remaining textual content is lemmatized.

\subsection{Feature engineering module}
\label{sec:feat_eng}

Regarding content-based features, we focus on stylistic data such as readability score, reading time, and the number of complex words in the post. Emotion and polarity data are also considered. Textual features are obtained as char and word $n$-grams. Table \ref{tab:features} details the features engineered. Some require \textit{ad hoc} processing, as explained in Section \ref{sec:feat_eng_results}. 

\begin{table}[!htbp]
\centering
\caption{\label{tab:features} Features engineered by data type.}
\begin{tabular}{llll}
\toprule
\bf Type & \bf ID & \bf Name & \bf Description \\\hline

\multirow{1}{*}{Boolean} & 1 & Emotion & \begin{tabular}[c]{@{}p{6.5cm}@{}} Indicates if the content expresses anger, fear, happiness, sadness, or surprise.\end{tabular}\\\midrule

\multirow{27}{*}{Numerical} & 2 & Complex word counter & \begin{tabular}[c]{@{}p{6.5cm}@{}} Number of complex words in the content.\end{tabular}\\

& 3 & \textsc{fre} & \begin{tabular}[c]{@{}p{6.5cm}@{}} Flesch reading ease score to assess readability based on the number of words and the number of syllables per word.\end{tabular}\\

& 4 & Hashtag counter & \begin{tabular}[c]{@{}p{6.5cm}@{}} Number of hashtags in the content.\end{tabular}\\

& 5 & \textsc{mer} & \begin{tabular}[c]{@{}p{6.5cm}@{}} McAlpine \textsc{eflaw} Readability score considering the number of words and mini-words (up to 3 characters) and the number of sentences.\end{tabular}\\

& 6 & Neg. num. & \begin{tabular}[c]{@{}p{6.5cm}@{}} Amount of negative numerical values referring to a stock or market.\end{tabular}\\

& 7 & Neg. per. & \begin{tabular}[c]{@{}p{6.5cm}@{}} Amount of negative percentages.\end{tabular}\\

& 8 & Polarity & \begin{tabular}[c]{@{}p{6.5cm}@{}} Indicates the sentiment of the content, negative (-1), neutral (0), or positive (+1).\end{tabular}\\

& 9 & \textsc{pos} distributions & \begin{tabular}[c]{@{}p{6.5cm}@{}} 
Percentages of adjectives, adverbs, auxiliary elements, determiners, nouns, pronouns, and punctuation marks.\end{tabular}\\

& 10 & Pos. num. & \begin{tabular}[c]{@{}p{6.5cm}@{}} Amount of positive numerical values referring to a stock or market.\end{tabular}\\
 
& 11 & Pos. per. & \begin{tabular}[c]{@{}p{6.5cm}@{}} Amount of positive percentages.\end{tabular}\\

& 12 & Reading time & \begin{tabular}[c]{@{}p{6.5cm}@{}} Estimated time needed to read the content, 14.69 ms per character.\end{tabular}\\

& 13 & Word count & \begin{tabular}[c]{@{}p{6.5cm}@{}} Number of words in the content.\end{tabular}\\\midrule

\multirow{4}{*}{Textual} & 14 & Char-grams & \begin{tabular}[c]{@{}p{6.5cm}@{}} List of char $n$-grams in the content.\end{tabular}\\

& 15 & Word-grams & \begin{tabular}[c]{@{}p{6.5cm}@{}} List of word $n$-grams in the content.\end{tabular}\\

& 16 & Word tokens & \begin{tabular}[c]{@{}p{6.5cm}@{}} List of char $n$-grams only from text inside word boundaries.\end{tabular}\\

\bottomrule
\end{tabular}
\end{table}

\subsection{Classification module}
\label{sec:classification}

The hybrid classification module combines categorization based on term detection and \textsc{ml} prediction. As forecast categories, we consider short-term drops, short-term rises, and `other,' a third category for any other post mentioning stock values. As for the short-term time range, we consider up to four weeks. For term-based detection, a term lexicon is defined for each target category. The system considers that a post belongs to a specific category if it contains at least one term in the corresponding lexicon\footnote{These lexica are composed of unique terms.}, generated in practice as described in Section \ref{sec:classification_results}. This prior term-based detection increases the system's efficiency, leaving the remaining posts to be classified using supervised \textsc{ml} models trained with the features detailed in Section \ref{sec:feat_eng}. The \textsc{ml} algorithms selected, based on their satisfactory performance in the literature \citep{Huang2021,Ryans2021,Alzanin2022,Gonzalez2022,Hudon2022,Jalal2022,Raheja2022,Bengesi2023,Darad2023}, are (\textit{i}) Linear Support Vector Machines (\textsc{lsvm}), (\textit{ii}) Decision Trees (\textsc{dt}), (\textit{iii}) k-Nearest Neighbors (\textsc{knn}), (\textit{iv}) Naive Bayes (\textsc{nb}), (\textit{v}) Random Forest (\textsc{rf}), and (\textit{vi}) Gradient Boosting (\textsc{gb}).

\subsection{Quality assessment module}
\label{sec:qassessment}

This module computes the quality of the predictions according to the category they were classified into. It compares the predicted forecast with the actual evolution of the stock values, identified with \$cashtags and financial tickers. The final quality rank depends on the number of successful predictions. Only posts mentioning stocks or markets are considered for quality evaluation.

\subsection{Explainability module}
\label{sec:explainability}

\begin{figure}[!htpb]
 \centering
 \includegraphics[scale=0.155]{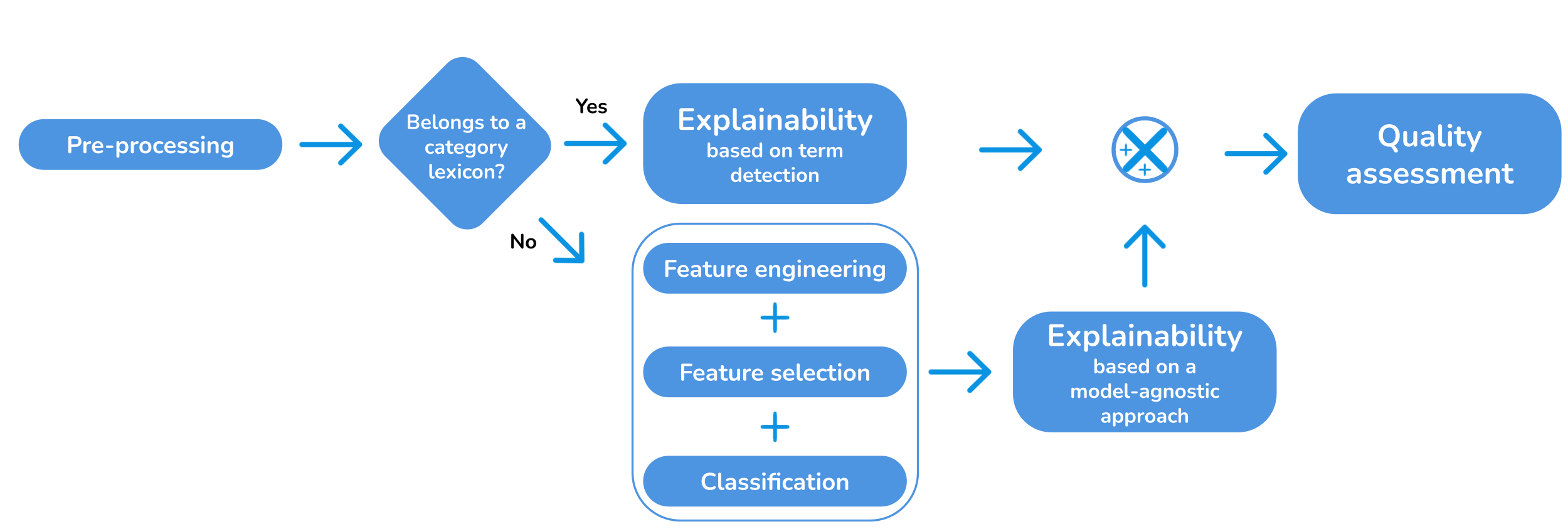}
 \caption{System's block diagram.}
 \label{fig:diagram}
\end{figure}

In order to foster the trustworthiness of the automatic analysis, this module presents explanations in natural language based on relevant terms or features. In the case of posts directly classified by term-based detection, the corresponding terms are included in the explanations. The system employs a model-agnostic approach for posts classified by the \textsc{ml} models to extract all the relevant features as explained in Section \ref{sec:explainability_results}. The explanations are presented in natural language using templates. Finally, Figure \ref{fig:diagram} shows the block diagram of the system from the pre-processing stage to the final quality assessment.

\section{Evaluation and discussion}
\label{sec:results}

This section presents the experimental data set, the implementations of the modules, and the results obtained. The experiments were performed on a computer with the following specifications:

\begin{itemize}
 \item \textbf{Operating System}. Ubuntu 20.04.3 \textsc{lts} 64 bits
 \item \textbf{Processor}. \textsc{amd}\textsuperscript\textregistered\ Ryzen 7 3800X 3.9 GHz
 \item \textbf{RAM}. 32 \textsc{gb} \textsc{ddr4} 
 \item \textbf{Disk}. 1 \textsc{tb} \textsc{ssd}
\end{itemize}

\subsection{Experimental data set}
\label{sec:dataset}

The experimental data set contains \num{15000} tweets published by 12 Spanish trading advisors (\num{1000} posts per advisor) between January 16, 2017, and June 5, 2023. Following a prior screening using ChatGPT, these posts were manually annotated by financial experts into three categories: (\textit{i}) short-term drop, (\textit{ii}) short-term rise, and (\textit{iii}) other, as previously described. The profiles of the creators were selected to ensure diversity in Twitter statistics (\textit{e.g.}, number of followers, posts per day, etc.). Table \ref{tab:adviser_distribution} shows the distribution of the samples in the experimental data set by category and advisor, while Table \ref{tab:category_distribution} details the global distribution of samples by category. 

\begin{table}[!htpb]\centering
\caption{Distribution of the samples by advisor and category.}
\label{tab:adviser_distribution}
\begin{tabular}{cccc}\toprule
\textbf{Ad.} & \textbf{Short-term drop (\%)} & \textbf{Short-term rise (\%)} & \textbf{Other (\%)} \\
\midrule
1 & 3.55 & 31.81 & 64.64 \\
2 & 17.64 & 42.89 & 39.48 \\
3 & 10.26 & 27.97 & 61.77 \\
4 & 13.81 & 26.23 & 59.96 \\
5 & 9.70 & 39.19 & 51.11 \\
6 & 5.00 & 77.70 & 17.30 \\
7 & 16.35 & 29.49 & 54.16 \\
8 & 5.94 & 30.38 & 63.68 \\
9 & 10.09 & 23.41 & 66.50 \\
10 & 7.34 & 34.31 & 58.35 \\
11 & 35.43 & 27.02 & 37.55 \\
12 & 11.42 & 23.75 & 64.83 \\
\bottomrule
\end{tabular}
\end{table}

\begin{table}[!htpb]\centering
\caption{Global distribution of samples by category.}
\label{tab:category_distribution}
\begin{tabular}{lc}\toprule
\textbf{Category} & \multicolumn{1}{l}{\textbf{\% samples}} \\
\midrule
Short-term drop & 12.20 \\
Short-term rise & 34.53 \\
Other & 53.26 \\ 
\bottomrule
\end{tabular}
\end{table}

\subsection{Pre-processing module}
\label{sec:preprocessing_results}

\$cashtags, images, links, and special characters were identified using the regular expressions in Listing \ref{lst:regular_expressions}.

\begin{lstlisting}[frame=single,caption={Regular expressions applied to detect \$cashtags, images, links, and special characters.}, label={lst:regular_expressions},emphstyle=\textbf,escapechar=ä]
ä\textbf{\$casthtags}ä = r'\$[a-zA-Z0-9=][a-zA-Z][a-zA-Z0-9=]+'
r'\^[a-zA-Z0-9=][a-zA-Z][a-zA-Z0-9=]+'
ä\textbf{Images/links}ä = r"(?:(pic.|http|www|\w+)?\:(//)*)\S+"
ä\textbf{Special char.}ä = r"(\*|\[|\]|=|\(|\)|\$|\"|\}|\{|\||\+|&|ä\texteuroä|ä\poundsä|/|ä\textdegreeä)+"

\end{lstlisting}

Other financial tickers not expressed in \$cashtag format (\textit{i.e.}, lacking the ``\$'' symbol or written as hashtags) were detected using regular expressions combined with dictionaries from the main American\footnote{Available at \url{https://www.sec.gov/file/company-tickers}, June 2024.} and European\footnote{Available at \url{https://www.cboe.com/europe/equities/market_statistics/symbols_traded}, June 2024.} stock markets. Moreover, stop words were removed using the Spanish \textsc{nltk} stop word list\footnote{Available at \url{https://www.nltk.org}, June 2024.} while Emoticons by applying the \textsc{nfkd} Unicode normalization form\footnote{Available at \url{https://unicode.org/reports/tr15}, June 2024.}. Finally, posts were lemmatized using the {\tt es\_core\_news\_md} core model of the \texttt{spaCy} Python library\footnote{Available at \url{https://spacy.io/models/es}, June 2024.}.

\subsection{Feature engineering module}
\label{sec:feat_eng_results}

Emotion and polarity (features 1 and 8 in Table \ref{tab:features}) were calculated with the \texttt{text2emotion}\footnote{Available at \url{https://pypi.org/project/text2emotion}, June 2024.} and \texttt{textblob} tools\footnote{Available at \url{https://spacy.io/universe/project/spacy-textblob}, June 2024.} from the \texttt{spaCy} Python library, respectively.

Stylistic features related to complex word count (feature 2), Flesch reading ease (feature 3), McAlpine \textsc{eflaw} Readability (feature 5), and reading time (feature 12) were obtained with the \texttt{textstat}\footnote{Available at \url{https://pypi.org/project/textstat}, June 2024.\label{fn:textstat}} Python library.

The distribution of \textsc{pos} tags in the posts (feature 9) was computed using the \texttt{textdescriptives}\footnote{Available at \url{https://pypi.org/project/textdescriptives}, June 2024.} Python library.

To compute the textual features, that is, char and word $n$-grams and word tokens (features 14, 15, and 16), the system employs the \texttt{CountVectorizer} function\footnote{Available at \url{https://scikit-learn.org/stable/modules/generated/sklearn.feature_extraction.text.CountVectorizer.html}, June 2024.\label{fn:cv}} from the \texttt{scikit-learn} Python library. Listings \ref{lst:param_c}-\ref{lst:param_wb} shows the optimal parameters in bold. Even though the ranges provided were established based on preliminary coarse experiments, the best values were identified using the \texttt{GridSearchCV}\footnote{Available at \url{https://scikit-learn.org/stable/modules/generated/sklearn.model_selection.GridSearchCV.html}, June 2024.\label{fn:gscv}} function from the \texttt{scikit-learn} Python library, as commonly used in the literature \citep{Evans2021}.

The rest of the features in Table \ref{tab:features} were computed using regular expressions or standard procedures in Python, as in the case of the hashtag counter (feature 4\footnote{r'\#[a-zA-Z0-9]+'\$.}), and the negative and positive number counters (features 6, 7, 10 and 11\footnote{r'-?[0-9.,]+[\$€\%]'.}). After content tokenization, feature 13 is engineered using the Python \texttt{len} function.

\begin{lstlisting}[frame=single,caption={Parameter selection for char-grams (best values in bold).}, label={lst:param_c},emphstyle=\textbf,escapechar=ä]
max_df = [0.1, 0.12, 0.14, 0.16, 0.18, 0.2, 0.22, 0.24, 0.26, 
0.28, 0.3, 0.32, 0.34, 0.36, 0.38, ä\textbf{0.4}ä],
min_df = [0.01, 0.02, ä\textbf{0.03}ä, 0.04, 0.05],
ngram_range = [ä\textbf{(3, 4)}ä, (3, 5), (3, 6), (3, 7), (4, 5), (4, 6), 
(5, 6)],
max_features = [100, 1000, ä\textbf{None}ä]
\end{lstlisting}

\begin{lstlisting}[frame=single,caption={Parameter selection for word-grams (best values in bold).}, label={lst:param_w},emphstyle=\textbf,escapechar=ä]
max_df = [0.17, 0.20, ä\textbf{0.23}ä, 0.26, 0.29, 0.32, 0.35, 0.38, 0.41, 
0.44, 0.47, 0.5],
min_df = [ä\textbf{0.01}ä, 0.02, 0.03, 0.04, 0.05],
ngram_range = [ä\textbf{(1, 1)}ä, (1, 2), (1, 3), (2, 2), (2, 3)],
max_features = [100, 1000, ä\textbf{None}ä]
\end{lstlisting}

\begin{lstlisting}[frame=single,caption={Parameter selection for word tokens (best values in bold).}, label={lst:param_wb},emphstyle=\textbf,escapechar=ä]
max_df = [0.17, 0.20, 0.23, ä\textbf{0.26}ä, 0.29, 0.32, 0.35, 0.38, 0.41, 
0.44, 0.47, 0.5],
min_df = [0.01, ä\textbf{0.02}ä, 0.03, 0.04, 0.05],
ngram_range = [(3, 4), ä\textbf{(3, 5)}ä, (3, 6), (3, 7), (4, 5), (4, 6), 
(5, 6)],
max_features = [100, 1000, ä\textbf{None}ä]
\end{lstlisting}

\subsection{Classification module}
\label{sec:classification_results}

The representative terms for each category lexicon were created using \texttt{CountVectorizer}\footref{fn:cv}. For each category, we selected the \SI{10}{\percent} most frequent unique terms (\textit{i.e.}, those which only appear in posts of a given category). If one post contained at least one term from the list of a category, it was assigned to that category. For the rest of the posts, the implementations for \textsc{ml} classification algorithms used were: 

\begin{itemize}
 \item \textsc{lsvm}. \texttt{LinearSVC} implementation\footnote{Available at \url{https://scikit-learn.org/stable/modules/generated/sklearn.svm.LinearSVC.html}, June 2024.} from the \texttt{scikit-learn} Python library.

 \item \textsc{dt}. \texttt{DecisionTreeClassifier} algorithm \footnote{Available at \url{https://scikit-learn.org/stable/modules/generated/sklearn.tree.DecisionTreeClassifier.html}, June 2024.} from the \texttt{scikit-learn} Python library.

 \item \textsc{knn}. \texttt{KNeighborsClassifier} implementation\footnote{Available at \url{https://scikit-learn.org/stable/modules/generated/sklearn.neighbors.KNeighborsClassifier.html}, June 2024.} from the \texttt{scikit-learn} Python library.
 
 \item \textsc{nb}. Two configurations, Complement Naive Bayes (\textsc{cnb}) and Multinomial Naive Bayes (\textsc{mnb}). The corresponding implementations were \texttt{ComplementNB}\footnote{Available at \url{https://scikit-learn.org/stable/modules/generated/sklearn.naive_bayes.ComplementNB.html}, June 2024.} and \texttt{MultinomialNB}\footnote{Available at \url{https://scikit-learn.org/stable/modules/generated/sklearn.naive_bayes.MultinomialNB.html}, June 2024.}, both from the \texttt{scikit-learn} Python library.

 \item \textsc{rf}. \texttt{RandomForestClassifier} function\footnote{Available at \url{https://scikit-learn.org/stable/modules/generated/sklearn.ensemble.RandomForestClassifier.html}, June 2024.} from the \texttt{scikit-learn} Python library.
 
 \item \textsc{gb}. \texttt{GradientBoosting} \texttt{Classifier} function\footnote{Available at \url{https://scikit-learn.org/stable/modules/generated/sklearn.ensemble.GradientBoostingClassifier.html}, June 2024.} from the \texttt{scikit-learn} Python library. 
 
\end{itemize}

We estimated the optimal hyperparameters\footnote{Note that we refer to parameters when generating features, while hyperparameters are those applied in \textsc{ml} models.} for these models using \texttt{Grid} \texttt{SearchCV}\footref{fn:gscv}, similarly to the feature engineering module. Listings \ref{lst:hyperparam_lsvm}-\ref{lst:hyperparam_gb} contain the ranges of values tested. In the case of \textsc{knn}, the only valid setting for the \texttt{metric} parameter was Minkowski, as the input matrix was not square.

Term-based detection allowed predicting \SI{62.99}{\percent} of the samples in the experimental data set, considerably saving computing resources and time.
For the subsequent evaluation with \textsc{ml} classification algorithms, we applied $10$-fold cross-validation \citep{Aoumeur2023}, implemented using the \texttt{StratifiedKFold}\footnote{Available at \url{https://scikit-learn.org/stable/modules/generated/sklearn.feature_selection.SelectPercentile.html}, June 2024.} function from the \texttt{scikit-learn} Python library. 

\begin{lstlisting}[frame=single,caption={Hyperparameter selection for \textsc{lsvm} (best values in bold).}, label={lst:hyperparam_lsvm},emphstyle=\textbf,escapechar=ä]
class_weight = [ä\textbf{None}ä, balanced],
loss = [ä\textbf{hinge}ä, squared_hinge],
max_iter = [500, ä\textbf{1000}ä, 2000],
multi_class = [ä\textbf{ovr}ä],
tol = [0.000001, 0.00001, 0.0001, 0.001, ä\textbf{0.01}ä],
penalty = [ä\textbf{l2}ä],
dual = [ä\textbf{True}ä],
C = [0.0001, 0.001, 0.01, ä\textbf{0.1}ä, 1, 10]
\end{lstlisting}

\begin{lstlisting}[frame=single,caption={Hyperparameter selection for \textsc{dt} (best values in bold).}, label={lst:hyperparam_dt},emphstyle=\textbf,escapechar=ä]
criterion = [ä\textbf{gini}ä, entropy],
splitter = [ä\textbf{best}ä, random],
class_weight = [ä\textbf{None}ä, balanced],
max_features = [None, ä\textbf{sqrt}ä, log2],
max_depth = [5, 10, 50, ä\textbf{100}ä],
min_samples_split = [0.0001, 0.001, ä\textbf{0.01}ä, 0.1, 1],
min_samples_leaf = [0.0001, 0.001, 0.01, 0.1, ä\textbf{1}ä]
\end{lstlisting}

\begin{lstlisting}[frame=single,caption={Hyperparameter selection for \textsc{knn} (best values in bold).}, label={lst:hyperparam_knn},emphstyle=\textbf,escapechar=ä]
n_neighbors = [5, ä\textbf{10}ä, 25],
weights = [uniform, ä\textbf{distance}ä],
algorithm = [auto, ball_tree, ä\textbf{kd\_tree}ä, brute],
leaf_size = [5, ä\textbf{10}ä, 15, 25, 30, 50],
p = [ä\textbf{1}ä, 2],
metric = [ä\textbf{minkowski}ä]
\end{lstlisting}

\begin{lstlisting}[frame=single,caption={Hyperparameter selection for \textsc{cnb} (best values in bold).}, label={lst:hyperparam_cnb},emphstyle=\textbf,escapechar=ä]
alpha = [0, ä\textbf{0.25}ä, 0.5, 0.75, 1],
fit_prior = [ä\textbf{True}ä, False],
norm = [True, ä\textbf{False}ä]
\end{lstlisting}

\begin{lstlisting}[frame=single,caption={Hyperparameter selection for \textsc{mnb} (best values in bold).}, label={lst:hyperparam_mnb},emphstyle=\textbf,escapechar=ä]
alpha = [0, ä\textbf{0.25}ä, 0.5, 0.75, 1],
fit_prior = [ä\textbf{True}ä, False]
\end{lstlisting}

\begin{lstlisting}[frame=single,caption={Hyperparameter selection for \textsc{rf} (best values in bold).}, label={lst:hyperparam_rf},emphstyle=\textbf,escapechar=ä]
n_estimators = [50, 100, ä\textbf{250}ä, 500, 1000],
criterion = [ä\textbf{gini}ä, entropy],
class_weight = [ä\textbf{None}ä, balanced],
max_features = [None, ä\textbf{sqrt}ä, log2],
max_depth = [5, 10, ä\textbf{50}ä, 100],
min_samples_split = [0.0001, ä\textbf{0.001}ä, 0.01, 0.1, 1],
min_samples_leaf = [ä\textbf{0.0001}ä, 0.001, 0.01, 0.1, 1]
\end{lstlisting}

\begin{lstlisting}[frame=single,caption={Hyperparameter selection for \textsc{gb} (best values in bold).}, label={lst:hyperparam_gb},emphstyle=\textbf,escapechar=ä]
learning_rate = [0.001, 0.01, 0.1, ä\textbf{1}ä],
n_estimators = [10, 100, ä\textbf{1000}ä],
max_depth = [5, 100, ä\textbf{100}ä, 1000],
min_samples_split = [0.0001, 0.001, 0.01, 0.1, ä\textbf{1}ä],
min_samples_leaf = [0.0001, 0.001, 0.01, 0.1, ä\textbf{1}ä],
max_features = [ä\textbf{sqrt}ä, None],
subsample = [0.1, 0.5, ä\textbf{1}ä],
tol = [0.000001, 0.00001, 0.0001, ä\textbf{0.001}ä],
\end{lstlisting}

Table \ref{tab:classification_results} shows the combined results of term-based and \textsc{ml} classifications. The best values for each measurement are expressed in bold. Note that the best model varies depending on the metric analyzed (i.e., precision, recall, and $F$-measure) and the target category (i.e., drop and rise). In the particular application of this work, it is necessary to maximize precision by assuming that the quality assessment module receives enough correctly predicted samples.

In light of the results obtained, we can conclude that the most promising models are \textsc{knn}, \textsc{mnb}, \textsc{rf}, and \textsc{gb} with the best values superior to \SI{80}{\percent} for all metrics. However, other aspects must also be considered, such as computing time. In this regard,  \textsc{knn} and \textsc{gb} are highly demanding. Lastly, taking into account the interpretable nature of the \textsc{rf} model in favor of explainability, it was selected. On the other hand, Table \ref{tab:user_results} shows the performance of the final hybrid classification module.

\begin{table}[!htbp]
\begin{center}
\begin{minipage}{\textwidth}
\caption{Classification metrics (lexica + \textsc{ml}).}
\label{tab:classification_results}
\begin{tabular*}{\textwidth}{@{\extracolsep{\fill}}lcccccccc@{\extracolsep{\fill}}}
\toprule
\multirow{2}{*}{\textbf{Model}} & \multicolumn{2}{@{}c@{}}{\textbf{Precision (\%)}} & \multicolumn{2}{@{}c@{}}{\textbf{Recall (\%)}} & \multicolumn{2}{@{}c@{}}{\textbf{F1 (\%)}} & \multicolumn{2}{@{}c@{}}{\textbf{Time (s)}} \\ \cmidrule{2-3}\cmidrule{4-5}\cmidrule{6-7}\cmidrule{8-9}
& Drop & Rise & Drop & Rise & Drop & Rise & Fit & Predict \\ \midrule
\textsc{lsvm} & 93.51 & 91.29 & 71.22 & 84.47 & 80.86 & 87.75 & 55.70 & 0.0057 \\
\textsc{dt} & 91.08 & 88.76 & 68.75 & 81.82 & 78.36 & 85.15 & 0.03 & 0.0006 \\
\textsc{knn} & 91.32 & \bf 91.47 & 71.50 & 81.70 & 80.20 & 86.31 & 0.01 & 0.8525 \\
\textsc{mnb} & 72.13 & 90.75 & \bf 82.83 & 80.24 & 77.11 & 85.17 & 0.02 & 0.0005 \\
\textsc{cnb} & 80.83 & 89.08 & 77.34 & 82.40 & 79.05 & 85.61 & 0.02 & 0.0005 \\
\textsc{rf} & \bf 93.77 & 89.39 & 72.39 & 85.87 & 81.71 & 87.60 & 1.89 & 0.0508 \\
\textsc{gb} & 88.90 & 90.36 & 76.99 & \bf 88.25 & \bf 82.52 & \bf 89.29 & 39.73 & 0.1174\\
\bottomrule
\end{tabular*}
\end{minipage}
\end{center}
\end{table}

\begin{table}[!htbp]
\begin{center}
\begin{minipage}{\textwidth}
\caption{Classification metrics by advisor (lexica + \textsc{rf}).}
\label{tab:user_results}
\begin{tabular*}{\textwidth}{@{\extracolsep{\fill}}lccccccc@{\extracolsep{\fill}}}\toprule
\multirow{2}{*}{\textbf{Ad.}} & \multicolumn{3}{c}{\textbf{Samples}} & \multicolumn{2}{c}{\textbf{Precision (\%)}} & \multicolumn{2}{c}{\textbf{Recall (\%)}} \\ \cmidrule{2-4}\cmidrule{5-6}\cmidrule{7-8}
& Drop & Rise & Other & Drop & Rise & Drop & Rise \\\midrule
1 & 35 & 314 & 638 & 86.67 & 97.35 & 74.29 & 93.63 \\
2 & 176 & 428 & 394 & 97.33 & 90.00 & 82.95 & 94.63 \\
3 & 102 & 278 & 614 & 95.45 & 84.78 & 61.76 & 84.17 \\
4 & 138 & 262 & 599 & 92.00 & 84.54 & 50.00 & 66.79 \\
5 & 96 & 388 & 506 & 88.89 & 86.93 & 66.67 & 84.02 \\
6 & 50 & 777 & 173 & 100.00 & 93.67 & 76.00 & 98.97 \\
7 & 163 & 294 & 540 & 99.15 & 87.26 & 71.78 & 93.20 \\
8 & 59 & 302 & 633 & 95.12 & 86.27 & 66.10 & 58.28 \\
9 & 100 & 232 & 659 & 92.31 & 83.43 & 60.00 & 65.09 \\
10 & 73 & 341 & 580 & 96.00 & 93.00 & 65.75 & 81.82 \\
11 & 350 & 267 & 371 & 91.74 & 85.08 & 85.71 & 94.01 \\
12 & 114 & 237 & 647 & 91.30 & 87.55 & 73.68 & 86.08 \\
\bottomrule
\end{tabular*}
\end{minipage}
\end{center}
\end{table}

\subsection{Quality assessment module}
\label{sec:qassessment_results}

In order to assess the predictions, we obtained the actual market prices for the stocks mentioned in the posts. A valuable resource for this purpose is the \texttt{yfinance} Python library\footnote{Available at \url{https://pypi.org/project/yfinance}, June 2024.}, which retrieves information about stock prices from Yahoo\textsuperscript\textregistered finance \textsc{api}\footnote{Available at \url{https://finance.yahoo.com}, June 2024.}.

In this research, the posts were annotated as short-term forecasts or one type or another if they referred to a maximum time range of about four weeks. The latter is because, in short-term trading, the results of the predictions are checked weekly \citep{Weng2018,Shen2020,Duarte2021}. Our assessment considered a drop (/rise) for a particular stock if its minimum (/maximum) price during three work weeks from Monday to Friday was at least \SI{3}{\percent} lower (/higher)\footnote{Established based on experimental tests.} than the stock's closing price the day the post was published. Listing \ref{lst:example_classifier} shows examples of two correctly classified posts, one of each essential category.

\begin{lstlisting}[frame=single,caption={Example of two correctly classified posts.}, label={lst:example_classifier},emphstyle=\textbf,escapechar=ä]
ä\textbf{Short-term rise.}ä ä\textit{\$NNOX Está alcista en el corto plazo y mejoraría bastante si superase}ä
ä\textit{ los 14,73.}ä
$NNOX is bullish in the short term and will improve if it gets
over 14.73.

ä\textbf{Short-term drop.}ä ä\textit{\$ASML inmerso en un canal bajista de corto plazo}ä
$ASML lost on a downward channel in the short term.

\end{lstlisting}

\begin{figure}
\centering
\begin{subfigure}{.5\textwidth}
 \centering
 \includegraphics[width=.95\linewidth]{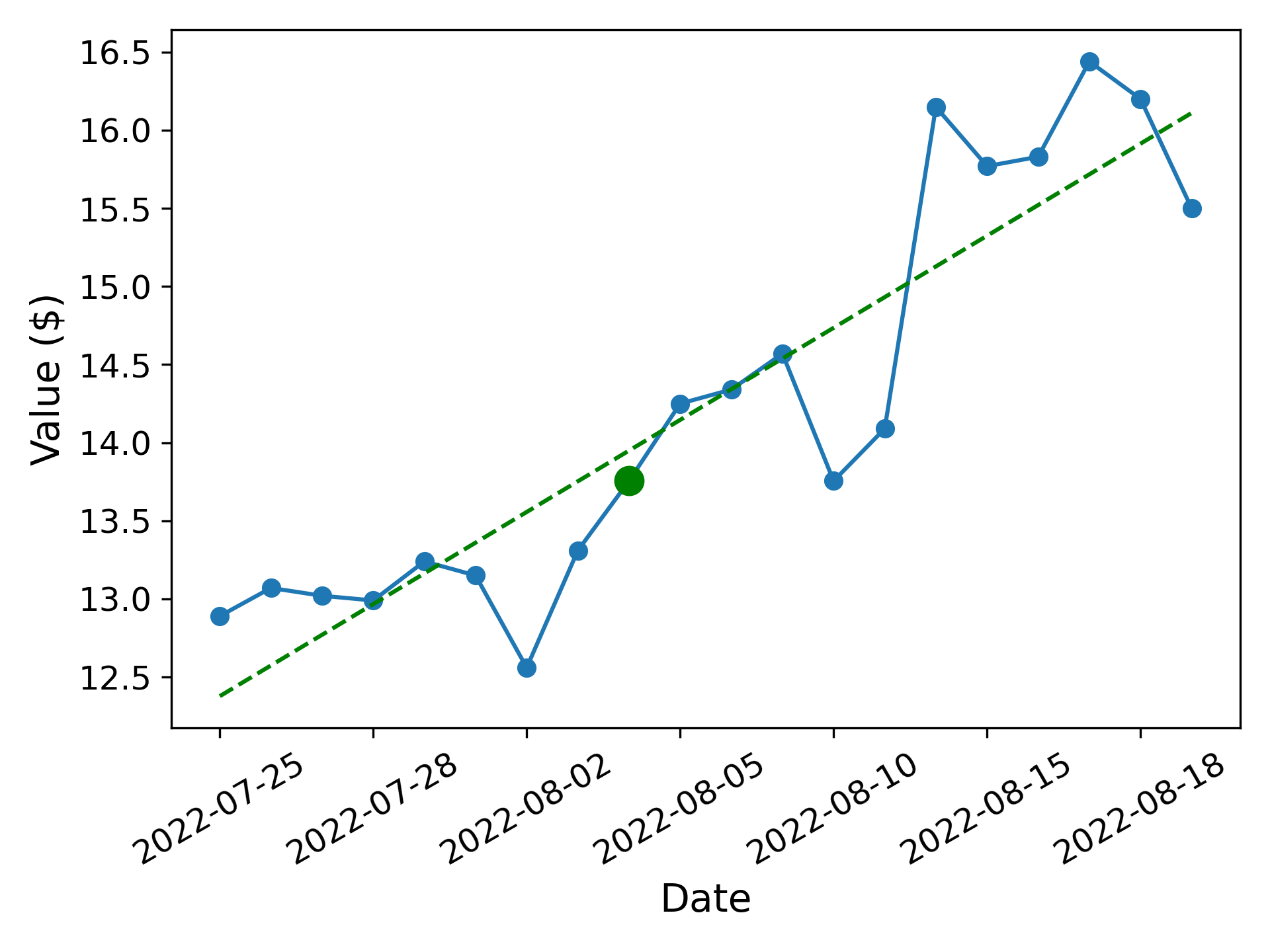}
 \caption{Nano-X Imaging Ltd stock price evolution.}
 \label{fig:nnox}
\end{subfigure}
\begin{subfigure}{.5\textwidth}
 \centering
 \includegraphics[width=.95\linewidth]{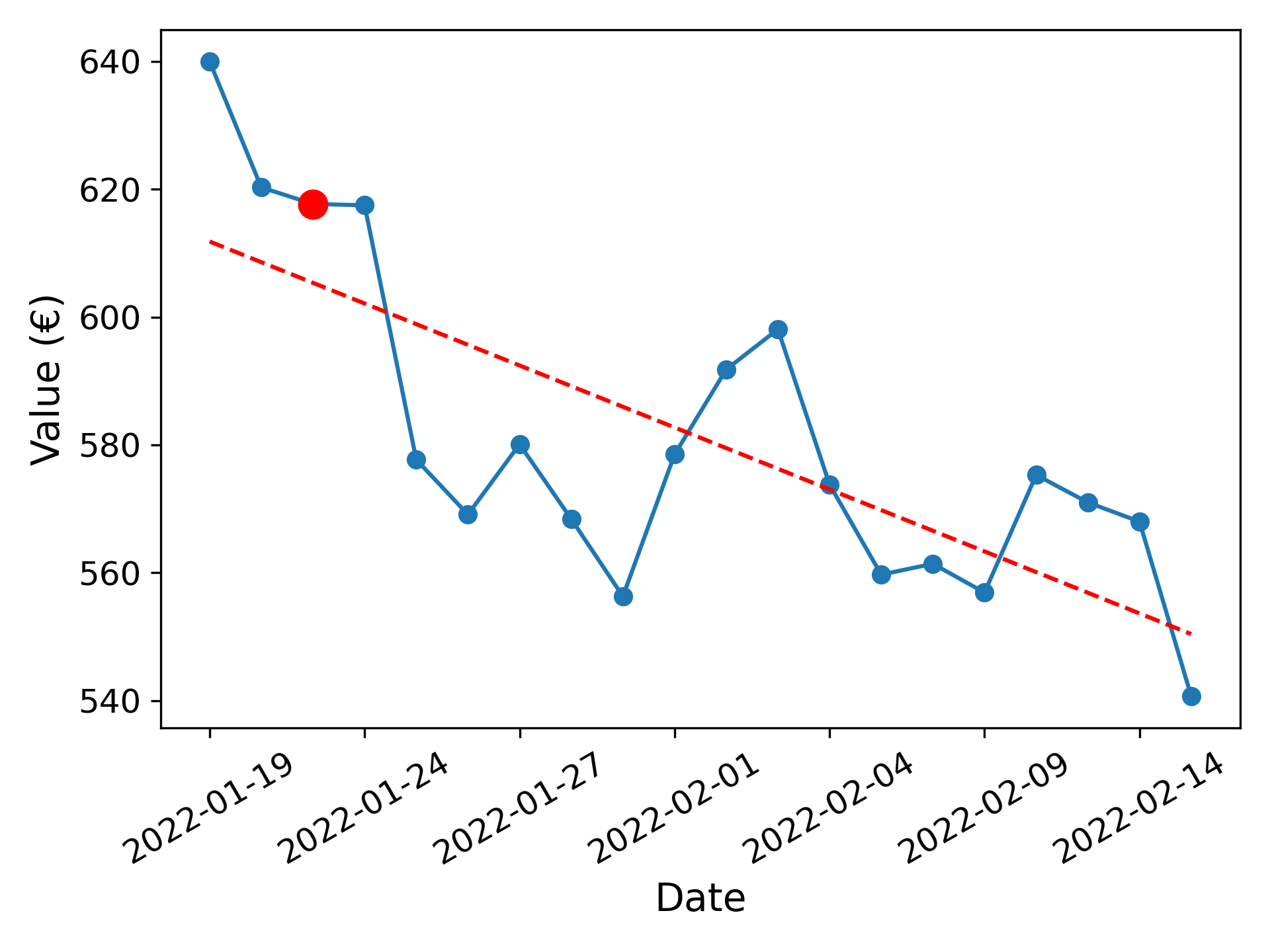}
 \caption{\textsc{asml} stock price evolution.}
 \label{fig:asml}
\end{subfigure}
\caption{Evolution of \$NNOX and \$ASML stocks.}
\label{fig:stocks}
\end{figure}

Figure \ref{fig:stocks} shows the evolution of these stocks' prices from the post's publication until the end of the 4-week window. Both plots clearly show that the predictions of the posts' forecasts are coherent with the evolution of the stock values. A colored dot marks the first day the 3\% difference was fulfilled. Therefore, both predictions were correct, and the respective quality rankings were updated.

Moreover, Table \ref{tab:qualities} shows the quality rankings of the predictions estimated per advisor as success probabilities, which are continuous variables, as desired. In light of the results, we can observe three different profiles: one more balanced, represented by advisors 2 and 12, with similar qualities for both drops and rises; another better in drops, like that of advisors 4, 5, and 7; and another in rises, like advisors 1 and 6.

Additionally, Table \ref{tab:social_metrics} shows the social network metrics (user-context metrics) of the advisors' accounts on June 5, 2023. The peak and average likes, retweets, and replies were calculated using the last 1,000 published posts.

Finally, we studied the correlations between the continuous quality rankings and the user-context metrics of the users' accounts, the number of followers, and the average and peak number of likes, retweets, and replies. This test offered insights into the reputation of the advisors in the social network and why the users preferred them. Table \ref{tab:quality_correlations} shows the correlations.

\begin{table}[!htbp]
\centering
\caption{Advisors' rankings.}
\label{tab:qualities}
\begin{tabular}{cccc}\toprule
\multirow{2}{*}{\textbf{Ad.}} & \multicolumn{3}{c}{\textbf{Prediction quality (\%)}} \\\cmidrule{2-4}
& Drop & Rise & Global \\\midrule
1 & 50.00 & 83.33 & 80.77 \\
2 & 76.04 & 78.00 & 77.27 \\
3 & 70.00 & 76.00 & 73.24 \\
4 & 79.63 & 66.10 & 69.61 \\
5 & 73.17 & 62.21 & 64.18 \\
6 & 60.61 & 67.49 & 67.36 \\
7 & 77.91 & 63.27 & 67.70 \\
8 & 80.95 & 75.86 & 76.25 \\
9 & 75.76 & 72.86 & 73.58 \\
10 & 90.91 & 84.34 & 85.26 \\
11 & 73.95 & 69.28 & 72.29 \\
12 & 64.37 & 63.37 & 63.91 \\
\bottomrule
\end{tabular}
\end{table}

\begin{table}[!htbp]
\begin{center}
\begin{minipage}{\textwidth}
\caption{Social network metrics of the advisors.}
\label{tab:social_metrics}
\begin{tabular*}{\textwidth}{@{\extracolsep{\fill}}clllllll@{\extracolsep{\fill}}}\toprule
\multirow{2}{*}{\textbf{Ad.}} & \multirow{2}{*}{\textbf{Followers}} & \multicolumn{2}{c}{\textbf{Retweets}} & \multicolumn{2}{c}{\textbf{Likes}} & \multicolumn{2}{c}{\textbf{Replies}} \\
\cmidrule{3-4}\cmidrule{5-6}\cmidrule{7-8}
& & Avg. & Max. & Avg. & Max. & Avg. & Max. \\\midrule
1 & 10800 & 6.50 & 66 & 5.95 & 404 & 0.19 & 22 \\
2 & 2657 & 40.57 & 12 & 3.74 & 52 & 0.73 & 74 \\
3 & 33933 & 176.11 & 1919 & 31.61 & 313 & 2.25 & 54 \\
4 & 18796 & 187.68 & 1188 & 18.74 & 246 & 1.18 & 26 \\
5 & 5218 & 245.02 & 56109 & 5.04 & 70 & 0.78 & 13 \\
6 & 891 & 1.32 & 8 & 0.78 & 11 & 0.08 & 2 \\
7 & 3609 & 4.39 & 9 & 4.36 & 34 & 0.34 & 12 \\
8 & 12062 & 37.29 & 1070 & 4.86 & 72 & 1.08 & 16 \\
9 & 8574 & 210.72 & 92565 & 10.55 & 254 & 0.84 & 38 \\
10 & 3465 & 183.45 & 34457 & 3.54 & 116 & 0.77 & 20 \\
11 & 4650 & 106.03 & 11 & 5.39 & 66 & 0.92 & 14 \\
12 & 3662 & 11.77 & 1049 & 2.90 & 160 & 0.70 & 43 \\
\bottomrule
\end{tabular*}
\end{minipage}
\end{center}
\end{table}

\begin{table}[!htbp]
\centering
\caption{Correlations between social network metrics and quality rankings.}
\label{tab:quality_correlations}
\begin{tabular}{lcccc}\toprule
\multicolumn{2}{c}{\multirow{2}{*}{\textbf{Social metric}}} & \multicolumn{3}{c}{\textbf{Correlations}} \\\cmidrule{3-5}
\multicolumn{2}{c}{} & Drop & Rise & Global\\\midrule
Followers & Num. & -0.029 & 0.184 & 0.091 \\
\multirow{2}{*}{Retweets} & Avg. & 0.473 & -0.018 & 0.031 \\
 & Max. & 0.264 & -0.007 & 0.021 \\
\multirow{2}{*}{Likes} & Avg. & 0.049 & 0.084 & 0.011 \\
 & Max. & -0.406 & 0.403 & 0.302 \\
\multirow{2}{*}{Replies} & Avg. & 0.331 & 0.107 & 0.056 \\
 & Max. & 0.030 & 0.284 & 0.180 \\
 \bottomrule
\end{tabular}
\end{table}

The strongest correlations in the table suggest that users are more fond of advisors who correctly predict rises and do not value those who predict drops. However, we can observe a relation between drop predictions with retweets and replies. The latter suggests that, even though drop predictions are less popular (as they are more challenging to convert into gains for inexpert investors) regarding likes, they attract the interest of investors who comment about them and share them with other users.

\subsection{Explainability module}
\label{sec:explainability_results}

A model-agnostic approach evaluates a model according to its performance regardless of its internal structure. The \textsc{lime}\footnote{Available at \url{https://lime-ml.readthedocs.io/en/latest}, June 2024.} (Local Interpretable Model-Agnostic Explanations) algorithm trains a local model that reproduces the performance of the original model and evaluates the impact of the presence (or absence) of a given input feature in the final classification result. 

From the resulting most relevant features, we explained in natural language the prediction outcome using the template in Listing \ref{lst:exp_template_en}, where \textless $tweet$\textgreater\ represents the original published post, \textless $category$\textgreater\ the prediction of the classification module and \textless $terms$\textgreater\ and \textless $features$\textgreater\ unique terms and the most relevant features involved in the classification, respectively.

As the textual features may include char-grams not corresponding to understandable words, we implemented the methodology in our previous research \citep{Gonzalez2023}, which approximates those char-grams to the extended most frequent words that contain them.

\begin{lstlisting}[frame=single,caption={Natural language explanation template (in English).}, label={lst:exp_template_en},emphstyle=\textbf,escapechar=ä]
The classification of the post ä\textcolor{blue}{\textless $tweet$\textgreater}ä as ä\textcolor{blue}{\textless $category$\textgreater}ä can be 
explained by the presence of these terms: ä\textcolor{blue}{\textless $terms$\textgreater}ä. 
Additionally, the process considered the following features: 
ä\textcolor{blue}{\textless $features$\textgreater}ä.
\end{lstlisting}

Listing \ref{lst:res_template} contains a real example of an explanation generated for a classified post.

\begin{lstlisting}[frame=single,caption={Explanation example.}, label={lst:res_template},emphstyle=\textbf,escapechar=ä]
The classification of the post ä``\textcolor{blue}{\#SANTANDER Bullish divergence}ä 
ä\textcolor{blue}{monitoring 4.54 euros at close}ä as ä\textcolor{blue}{short-term rise}ä can be explained by 
the presence of these terms: ä\textcolor{blue}{[`bullish', `divergence']}ä.
Additionally, the process considered the following features: 
ä\textcolor{blue}{[`Pos. num.']}ä.
\end{lstlisting}

\begin{figure}
\centering
\begin{subfigure}{.5\textwidth}
 \centering
 \includegraphics[width=.95\linewidth]{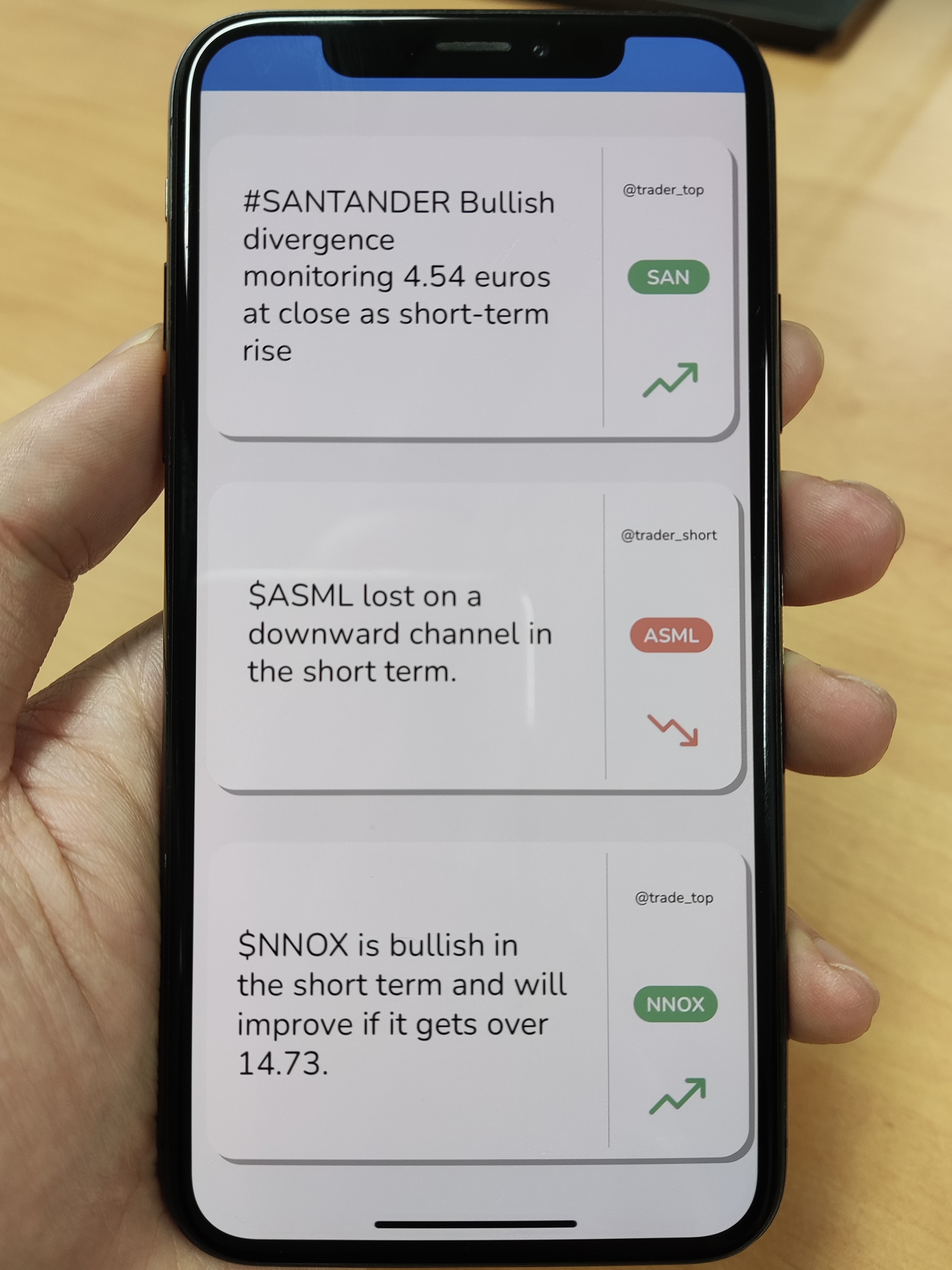}
 \caption{List of analyzed posts.}
 \label{fig:list_tweets}
\end{subfigure}%
\begin{subfigure}{.5\textwidth}
 \centering
 \includegraphics[width=.95\linewidth]{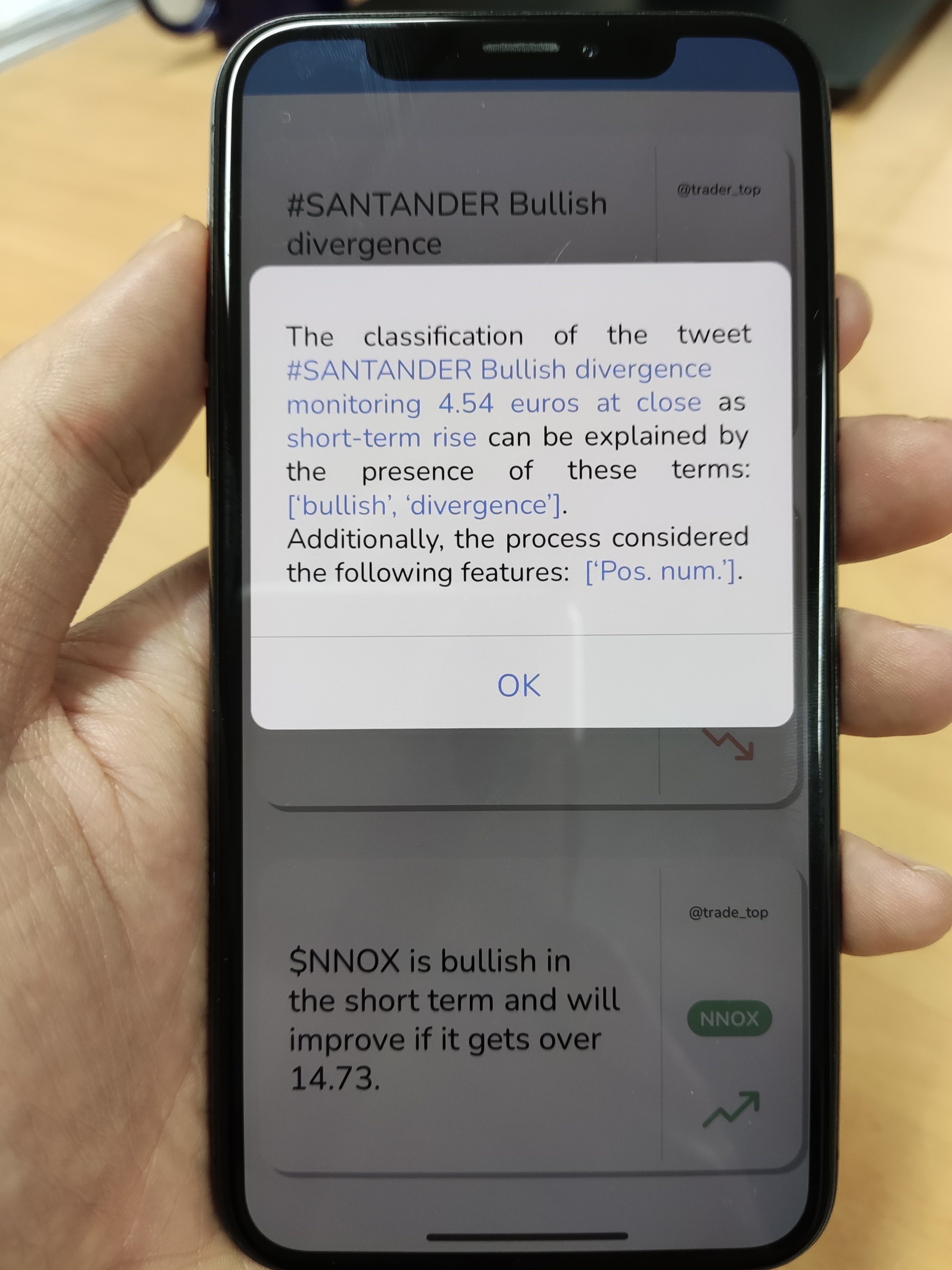}
 \caption{Selected post explanation.}
 \label{fig:xai_app}
\end{subfigure}
\caption{Mobile application.}
\label{fig:app}
\end{figure}

The post, classified as a short-term rise, contains unique terms \textit{bullish} and \textit{divergence}. Additionally, the model-agnostic methodology considered favorable prices (feature 10 in Table \ref{tab:features}) a relevant feature. This explanation in natural language is descriptive and understandable for a user without expert knowledge in finance or \textsc{ml}. Figure \ref{fig:app} shows the mobile application. In the picture on the left (a), the list of posts displayed includes information about the author, ticker, and short-term prediction. When the end user selects a particular post, a new window pops up (picture on the right (b)) to explain the prediction in natural language.

\subsection{Discussion and comparison with prior work}
\label{sec:comparison}

The credibility of social media in finance has already been studied in prior work. However, this research follows a more integral solution by comparing automatic classifications of posts' forecasts with the actual evolution of market values. The results are comparable or superior to those in the literature \citep{Kamkarhaghighi2016,Evans2021}.
Our system offers high classification performance and a quality ranking that we have compared with user-context social media metrics. The solution accurately classifies financial social media posts using a hybrid approach that combines term-based and \textsc{ml} techniques with average precision values exceeding \SI{90}{\percent} in the case of \textsc{rf}. It then automatically assesses user credibility from social media, providing a credibility ranking on a continuous scale. As valuable insights toward understanding human behavior related to stock market predictions, we have identified a positive correlation between drop predictions and retweets and replies and an inverse correlation between those predictions and the number of likes.

As a side contribution to explainability, we offer model-agnostic descriptions using natural language templates, which are legible and easily understandable. For example, the latter is more flexible than the approach by \citet{Huang2021}, which requires previous knowledge about the structure of the \textsc{dt} model and does not provide explanations in natural language.

\section{Conclusions}
\label{sec:conclusions}

Social media anonymity, fast-spreading, and low-effort information-sharing have raised credibility concerns about users regarded as experts. Since the amount of information users post is enormous, there is a need for automatic mechanisms to evaluate their reputation. Most current works on credibility assessment have mainly approached this task as a binary classification problem instead of ranking credibility as a continuous metric and do not compare their outcome with real-world data. Moreover, they do not have automatic explainability capabilities.

The main contribution of this research is a credibility assessment methodology combining \textsc{ml} and Natural Language Processing (\textsc{nlp}), applied to financial posts' forecasts, which are automatically classified and compared with the actual evolution of the market values. This allows ranking credibility to be a continuous metric of success probability. Moreover, the correlations between the credibility ranking and the social media metrics (\textit{i.e.}, the user context) produce insights related to the interest of the end-users regarding the posts' forecasts. Concerning explainability, the system generates descriptions of its decisions in natural language, which are legible and easily understandable, using a model-agnostic approach. Therefore, the classification can be entirely modular and interchangeable. This explainable system may allow non-expert investors to understand the classifier's decisions and quickly judge investment recommendations, thus empowering them with valuable financial knowledge on market strategies. The resulting application supports continuous real-time analysis of the posts (the prediction time with \textsc{rf} was 0.05 seconds).

Furthermore, given the fact that the underlying term-based and \textsc{ml} techniques can be applied to many \textsc{nlp} tasks (\textit{i.e.}, \textsc{sa}, detection of fake news, classification of medical or legal texts), the strategy in this work can be applied to any field with a strong presence in social media in which reputation assessment has value, such as recommendation platforms or online gaming, complementing traditional metrics like the past return rate (yield) and percentage of success.

In future work, we plan to take advantage of real-time data to train our models in streaming mode, introduce new approaches (\textit{e.g.}, reinforcement learning) to further enhance the system's adaptability, and analyze how new features influence the assessment, including the effect of the elapsed time between the post and the current date, which will be addressed with sliding windows. We will explore how to improve recall in our system by reducing the data drift problems in posts written by different authors. We will also study adaptive precision methods to select the safest investment recommendations and explore long-term stock market opportunities. Finally, we will consider the application to other fields, such as online gaming with sports bets.

\section*{CRediT authorship contribution statement}

\textbf{Silvia García-Méndez}: Conceptualization, Methodology, Software, Validation, Formal analysis, Investigation, Resources, Data Curation, Writing - Original Draft, Writing - Review \& Editing, Visualization, Supervision, Project administration, Funding acquisition. \textbf{Francisco de Arriba-Pérez}: Conceptualization, Methodology, Software, Validation, Formal analysis, Investigation, Resources, Data Curation, Writing - Original Draft, Writing - Review \& Editing, Visualization, Supervision, Project administration, Funding acquisition. \textbf{Jaime González-González}: Software, Validation, Resources, Data Curation, Writing – original draft. \textbf{Francisco J. González-Castaño}: Conceptualization, Methodology, Investigation, Writing - Review \& Editing, Supervision.

\section*{Declaration of competing interest}

The authors declare no competing financial interests or personal relationships that could influence the work reported in this paper.

\section*{Data availability}

Data are available on request from the authors.

\section*{Acknowledgments}

This work was partially supported by (\textit{i}) Xunta de Galicia grants ED481B-2021-118, ED481B-2022-093, and ED431C 2022/04, Spain; (\textit{ii}) Ministerio de Ciencia e Innovación grant TED2021-130824B-C21, Spain; and (\textit{iii}) University of Vigo/CISUG for open access charge.

\bibliography{bibliography}
\end{document}